\documentclass[aps,prd,nofootinbib,showpacs,12pt]{revtex4}
\usepackage{amsmath}

\begin{document}

\title{Boiling of nuclear liquid in core-collapse supernova explosions}

\author{Peter Fomin}\email{pfomin@bitp.kiev.ua}
\author{Dmytro Iakubovskyi}\email{yakubovskiy@bitp.kiev.ua}
\author{Yuri Shtanov}\email{shtanov@bitp.kiev.ua}
\affiliation{Bogolyubov Institute for Theoretical Physics, Kiev 03680, Ukraine}

\newcommand{\eq}[1]{\begin{equation} #1 \end{equation}}
\newcommand{\ml}[1]{\begin{multline} #1 \end{multline}}
\newcommand{\erf}{\mathop{\rm erf}\nolimits}
\newcommand{\tr}{\mathop{\rm tr}\nolimits}
\newcommand{\sgn}{\mathop{\rm sgn}\nolimits}
\newcommand{\p}{\partial}

\newcommand{\Sp}{\mathop{\rm Sp}\nolimits}

%%%%%%%%%%%%%%%%%%%%%%%%%%%%%%%%%%%%%%%%%%%%%%%%%%%%%%%%%%%%%%%%%%%%%%%

\begin{abstract}
We investigate the possibility of boiling instability of nuclear liquid in the
inner core of the proto-neutron star formed in the core collapse of a type~II
supernova. We derive a simple criterion for boiling to occur. Using this
criterion for one of best described equations of state of supernova matter, we
find that boiling is quite possible under the conditions realized inside the
proto-neutron star. We discuss consequences of this process such as the
increase of heat transfer rate and pressure in the boiling region. We expect
that taking this effect into account in the conventional neutrino-driven
delayed-shock mechanism of type~II supernova explosions can increase the
explosion energy and reduce the mass of the neutron-star remnant.
\end{abstract}


%%%%%%%%%%%%%%%%%%%%%%%%%%%%%%%%%%%%%%%%%%%%%%%%%%%%%%%%%%%%%%%%%%%%%%%

\maketitle

\vfill\eject

\section{ Introduction }\label{Introduction}

Detection of electron antineutrinos from SN~1987A (see \cite{H87,H88,Bi87})
confirmed the previous theoretical ideas about neutrinos playing crucial role
in the core collapse of type~II supernovae. According to well established
estimates, only about one percent of the gravitational binding energy (or $(1.4
\pm 0.4) \times 10^{51}$~erg for SN~1987A; see \cite{BP90}) is released in the
thermal and kinetic energies of the expanding ejecta; the remaining part is
carried away by different types of neutrinos \cite{Nad78}, which are known to
be trapped effectively within nuclear matter during the last stages of core
collapse (see \cite{ShaTyu} and references therein). Therefore, initially it
did not seem hard to explain the observed values of kinetic energy of the
expanding ejecta using neutrino-nucleon interaction as an effective channel of
energy transfer from the emitted neutrinos to nuclear matter \cite{CW66,A66}.
However, concrete realization of this mechanism of supernova explosions
encountered with great difficulties. Thus, it was realized that the prompt
shock wave, generated during the core collapse, fades away at the time scale of
several milliseconds, losing its energy on nuclear dissociation
\cite{M82,CBB84} and on the radiation of $\nu\bar{\nu}$-pairs \cite{B85}. As a
result, in simulations, the shock starts to move inward, pushed down by the
infalling matter, leading to the formation of a black hole rather than to
explosion.

Considerable efforts were made to remedy this situation. \citet{BW85} showed
that the shock wave could be revived about 100~msec after the core bounce if
neutrino luminosity were sufficiently enhanced. The way of such enhancement of
neutrino luminosity was opened after neutrino diffusion inside the center of
the proto-neutron star (PNS) was discovered \cite{BBC87} and found to be
convectively unstable (see \cite{Bu87,BJRK06} and references therein).
Nevertheless, numerical simulations of the neutrino-driven shock revival still
yield contradictory results. While explosions with kinetic energy up to $1.72
\times 10^{51}$\,erg are obtained in \cite{Sch06} ($0.5 \cdot 10^{51}$\,erg in
\cite{BJRK06}, $10^{50}$\,erg in \cite{KJH06}, $0.94 \cdot 10^{51}$ erg in
\cite{SJFK}), there is also a number of simulations where explosions are not
observed at all \cite{Lieb03,j04}.

This inconsistency between the observational data and the results
of core-collapse numerical simulations motivated some researches
to look for alternative mechanisms. The magnetorotational
mechanism, proposed by \citet{BK70}, can produce explosions with
energies up to $0.61 \times 10^{51}$\,erg at the time scale of
$\sim 0.5$~sec after the core bounce \cite{BKMA05,BKMA06}. The
acoustic mechanism, developed by \citet{BLD06}, also produces
explosion; although its energy is rather uncertain in their
numerical simulations, it develops for a time range of several
hundred milliseconds. Thus, even with new mechanisms and effects
taken into account, the resulting energy release is marginally
short of the observed values.

One can conclude that all mechanisms of core collapse require some additional
engine to provide the observable explosions.  Such a new engine is the subject
of the present paper.  By analogy with the work of \citet{BBC87}, we consider
an additional type of transport of nuclear matter in the core which is
different from diffusion and convection.  It occurs in the form of {\em
boiling\/}, i.e., first order transition between different phases of nuclear
matter.

We start with the observation that supernova matter at the inner core can exist
in several phases and their mixtures \cite{RPW83,HSY84,WK85,LFBHS,WSYE,WMSYE}.
Among them are the following (in the order of increasing density):
\begin{itemize}
\item
the spherical \textit{nuclei\/} phase
\item
the phase consisting of elongated nuclei (often called \textit{pasta\/})
\item
the \textit{slab-like\/} nuclei
\item
the phase with \textit{cylindrical holes\/}
\item
the phase with charged \textit{microscopic\/} bubbles, or \textit{cheesed\/}
phase (we will use this last term below not to confuse it with
\textit{macroscopic\/} bubbles of which we will speak later)
\item
the \textit{homogeneous\/} supernova matter
\end{itemize}

It is natural to expect that several phase transitions can occur during the
evolution of nuclear matter during as well as after the core collapse.
Numerical simulations of the nuclear-matter phase transitions in supernovae
were usually aimed at determining the thermodynamic properties at the
pre-bounce stage of collapse, and were needed to understand the development of
the prompt shock wave. Such transitions taking place in a rapidly changing
environment during collapse can be called ``short-term'' phase transitions. In
our paper, we discuss what we call ``long-term'' phase transitions, which occur
in nuclear matter after the bounce under the condition of relative mechanical
and \textit{local thermodynamic equilibrium\/}.  The main goal of this paper is
to examine the potential importance of such ``long-term'' phase transitions on
the dynamics of supernova explosion.

In the ``normal'' case of mechanical and local thermodynamic equilibrium,
different phases occupy the corresponding radial shells of the PNS, and this
spatial phase picture evolves continuously and relatively slowly in time.
However, if heating of the nuclear matter (which is mainly due to diffusion of
neutrinos) is sufficiently strong and inhomogeneous, it can lead to a condition
similar to that in an ordinary teakettle. Specifically, a bulk of particular
phase can overheat and become unstable with respect to phase transition, and
small bubbles of the lighter phase can spontaneously appear in its volume. The
bubbles will grow and, due to the Archimedean force, rise upwards. Henceforth,
such a process is called \textit{boiling\/} by analogy. In this paper, we argue
that this process can provide more efficient mechanism of heating the outer
parts of the PNS (compared with the neutrino-diffusion and convection
mechanisms) and generate additional pressure wave.

The paper is organized as follows. In Sec.~\ref{Transition}, we give the basic
thermodynamic description of the co-existence of various phases in supernova
matter after the bounce. In Sec.~\ref{Boiling}, we derive the criterion of
boiling to occur, and, in Sec.~\ref{Estimates}, we provide our numerical
estimates for the values involved in this criterion using the tabulated
equation of state from \cite{suraud85}. In Sec.~\ref{Model}, we consider a
simple model of the boiling mechanism and derive numerical estimates
characterizing the efficiency of this process. We summarize and discuss our
results in Sec.~\ref{Discussion}.

\section{Phase equilibrium}\label{Transition}

Just after the core bounce, the inward movement of the inner core significantly
decreases, and the prompt shock wave moves outwards losing its energy mostly on
nuclear dissociation and radiation of $\nu\bar{\nu}$-pairs.  The material
within and around the PNS approaches a state of mechanical equilibrium (while
the velocities of the convective motion are much smaller than the appropriate
first cosmic velocities). In contrast with the rapid collapse during the
pre-bounce stage of contraction, local thermal equilibrium is a good
approximation for the post-bounce stage.

Because of the radial density gradient, the PNS at this stage has onion-like
structure. The density of its inner core is higher than the nuclear saturation
density, and the homogeneous matter phase is thermodynamically preferred. In
the outer layers of the PNS, the density is much less than the nuclear
saturation density, and matter exists in the form of ordinary nuclei. A number
of intermediate phases can exist between these two shells. Adjacent phases are
separated by the surfaces of coexistence.  Here we derive a simple condition of
coexistence of phases applied to the situation under consideration.

We can consider a phase transition characterized by fixed pressure $p$,
temperature $T$, baryon number $B$, electric charge $C$ and lepton number $L$.
The conservation laws of the last three quantities imply the existence of the
corresponding chemical potentials: $\mu_{B}$, $\mu_{C}$ and $\mu_{L}$. We can
determine their values from the chemical potentials of the supernova matter
components (we suppose that the supernova matter consists only of neutrons,
protons, electrons and electron neutrinos): \eq{\mu_{n} = \mu_{B}\, , \quad
\mu_{p} = \mu_{B} + \mu_{C}\, , \quad \mu_{e} = \mu_{L} - \mu_{C}\, , \quad
\mu_{\nu} = \mu_{L}\, .} Since the number of particle species is greater than
the number of independent charges, we have one more relation on the chemical
potentials (the so-called beta-equilibrium condition): \eq{\mu_{p} + \mu_{e} =
\mu_{n} + \mu_{\nu}\, .\label{equilibrium}}

For fixed values of $p$ and $T$, our thermodynamic system tends to a state with
a minimum value of the Gibbs free energy (see, e.g., \cite{landau5}) \eq{\Phi =
\sum_{i}\mu_{i}N_{i} = \mu_{B}B + \mu_{C}C + \mu_{L}L \, .\label{fi}} Since we
are interested only in electrically neutral phases, we have \eq{C\equiv 0\, ,}
and the second term in (\ref{fi}) vanishes. The third term, $\mu_{L}L$, does
not change during the phase transition because the electron neutrinos interact
with nuclear matter very weakly.

Finally, the condition of equilibrium between phases 1 and 2 is \eq{\mu_{1n}
\left( p_{0},T_{0},\mu_{\nu 0} \right) = \mu_{2n} \left( p_{0},T_{0},\mu_{\nu
0} \right) \, ,\label{coex}} where the subscript ``{\small 0}'' marks the
values of the parameters right on the interface between the two phases.

\section{Criterion of boiling}\label{Boiling}

Local thermal equilibrium described in the previous section is continuously
disturbed by the process of diffusion of electron neutrinos away from the
central region of the PNS.  This causes inhomogeneous heating of the bulk of
nuclear matter.  If this heating is sufficiently strong, it can lead to an
overheat of a particular phase, which may result in its {\em boiling\/}
--- emergence of bubbles of the adjacent lighter phase in its bulk.
These bubbles will then raise and grow, effectively transferring heat and
lepton number. Boiling is a particular case of a non-equilibrium first order
phase transition.  In this section, we derive a necessary condition of boiling
in the form of a relation on the parameters of the supernova matter.

According to Eq.~(\ref{coex}), in the process of external heating, a heavier
phase 1 becomes metastable with respect to its transition to a lighter phase 2
if the condition is reached such that \eq{\mu_{1n}(p_{0}+\delta p, T_{0}+\delta
T, \mu_{\nu 0} + \delta \mu_{\nu})> \mu_{2n}(p_{0}+\delta p, T_{0}+\delta T,
\mu_{\nu 0} + \delta \mu_{\nu}) \, ,} where the positive increments of the
thermodynamic variables correspond to their radial gradients. From this, we
obtain the condition
\begin{eqnarray} \label{main}
\left[\left(\frac{\partial \mu_{2n}}{\partial T}\right)_{p,\mu_{\nu}} - \left(
\frac{\partial \mu_{1n}}{\partial T}\right)_{p,\mu_{\nu}}\right]\frac{dT}{dr}
&+& \left[\left(\frac{\partial \mu_{2n}}{\partial \mu_{\nu}}\right)_{p,T} -
\left(\frac{\partial \mu_{1n}}{\partial \mu_{\nu}}\right)_{p,T}\right]
\frac{d\mu_{\nu}}{dr} \nonumber \\ &+& \left[\left(\frac{\partial
\mu_{2n}}{\partial p}\right)_{T,\mu_{\nu}} - \left(\frac{\partial
\mu_{1n}}{\partial p}\right)_{T,\mu_{\nu}}\right]\frac{dp}{dr}>0 \, .
\end{eqnarray}
In the important particular case of hydrostatic equilibrium, we have
\eq{\frac{dp}{dr} = -\rho_{m}g\, ,} where $\rho_{m}$ is the mean density, and
$g$ is the local acceleration of free fall. Substituting this into
(\ref{main}), we obtain
\begin{eqnarray} \label{main1}
\left[\left(\frac{\partial \mu_{2n}}{\partial T}\right)_{p,\mu_{\nu}} - \left(
\frac{\partial \mu_{1n}}{\partial T}\right)_{p,\mu_{\nu}}\right]\frac{dT}{dr}
&+& \left[\left(\frac{\partial \mu_{2n}}{\partial \mu_{\nu}}\right)_{p,T} -
\left(\frac{\partial \mu_{1n}}{\partial \mu_{\nu}}\right)_{p,T}\right]
\frac{d\mu_{\nu}}{dr} \nonumber \\ &>& \left[\left(\frac{\partial
\mu_{2n}}{\partial p}\right)_{T,\mu_{\nu}} - \left(\frac{\partial
\mu_{1n}}{\partial p}\right)_{T,\mu_{\nu}}\right]\rho_{m}g \, .
\end{eqnarray}

\section{Numerical estimates}\label{Estimates}

It remains to check whether the condition of boiling derived in the previous
section can be realized in the usual supernova core-collapse.  For this
purpose, we estimate the values of the partial derivatives in (\ref{main1})
using the numerical simulations of the EoS of supernova matter given in
\cite{suraud85}.

The authors of \cite{suraud85} tabulate all essential parameters describing the
supernova matter in three phases: nuclei, cheesed phase and homogeneous matter.
In Table~\ref{input}, we show the parameters which are required in our
analysis. Because the required values of derivatives are not listed in
\cite{suraud85}, we try to obtain them from interpolation. Namely, we use the
finite differences: \eq{\mu_{n}(p_{2},T_{2},\mu_{\nu
2})-\mu_{n}(p_{1},T_{1},\mu_{\nu 1}) \approx \frac{\partial \mu_{n}}{\partial
p} (p_{2}-p_{1}) + \frac{\partial \mu_{n}}{\partial T} (T_{2}-T_{1}) +
\frac{\partial \mu_{n}}{\partial \mu_{\nu}} (\mu_{\nu 2}-\mu_{\nu 1})\, ,}
neglecting the higher derivatives. For each phase, we solve a system of three
equations for three unknown variables.  The numerical values obtained in such a
manner should be considered as estimates.  They are presented in
Table~\ref{results}.
\begin{table}
{
\begin{tabular}{|l|c|c|c|c|c|}
\hline
Phase & $n_{B}$, fm$^{-3}$ & $\mu_{n}$, MeV & $p$,
MeV\,$\cdot$\,fm$^{-3}$ & $T$, MeV & $\mu_{\nu}$, MeV\\
\hline \hline
Nuclei & 0.02 & $-2.230$ & 0.1807 & 4.00 & 81.613\\
\hline
 & 0.04 & $-1.953$ & 0.4556 & 4.88 & 105.363\\
\hline
 & 0.05 & $-1.837$ & 0.6111 & 5.17 & 114.092\\
\hline
 & 0.06 & $-1.818$ & 0.7738 & 5.44 & 121.699\\
\hline \hline
Cheesed & 0.05 & $-2.660$ & 0.5615 & 5.07 & 114.149\\
\hline
 & 0.07 & $-2.189$ & 0.9086 & 5.65 & 127.934\\
\hline
 & 0.09 & $-1.967$ & 1.2726 & 6.08 & 139.143\\
\hline
 & 0.10 & $-2.093$ & 1.4432 & 6.23 & 144.213\\
\hline \hline
Homogeneous & 0.10 & $-2.536$ & 1.4241 & 6.18 & 144.530\\
\hline
 & 0.11 & $-1.202$ & 1.7203 & 6.64 & 147.952\\
\hline
 & 0.12 & 0.154 & 2.2267 & 7.09 & 169.312\\
\hline
 & 0.16 & 7.525 & 4.227 & 8.84 & 188.917\\
 \hline
\end{tabular}
} \caption{The parameters from \cite{suraud85} used in the estimate of the
partial derivatives in (\ref{main1}).} \label{input}
\end{table}

\begin{table}
{
\begin{tabular}{|l|c|c|c|}
\hline Phase & $\left(\frac{\partial \mu_{n}}{\partial
p}\right)_{T,\mu_{\nu}}$, fm$^{3}$ & $\left(\frac{\partial \mu_{n}}{\partial
T}\right)_{p,\mu_{\nu}}$
& $\left(\frac{\partial \mu_{n}}{\partial \mu_{\nu}}\right)_{p,T}$\\
\hline \hline
Nuclei & $-1.353$ & $-2.626$ & 0.125\\
\hline
Cheesed & 0.666 & 5.182 & $-0.201$\\
\hline
Homogeneous & 2.87 & 1.289 & $-0.0317$\\
\hline
\end{tabular}
} \caption{The results for the partial derivatives.} \label{results}
\end{table}

To check whether the conditions of boiling are realized, we use the
conventional values of the other parameters. Numerical simulations (see, e.g.,
\cite{Lieb03}) give the following estimates: \eq{g \approx 1.0 \times 10^{14}\,
\mbox{cm\,$\cdot$\,sec$^{-2}$}\, , \quad \frac{d\mu_{\nu}}{dr}\approx
-(\mbox{10$-$20})\, \mbox{MeV\,$\cdot$\,km$^{-1}$} \, , \quad \frac{dT}{dr}<0\,
.\label{data}} Using this data together with Table~\ref{results}, we obtain the
following conditions for boiling:
\begin{itemize}
\item for the transition between the nuclear and cheesed phase
($\rho_{m} \approx 0.8 \times 10^{14}$\,g\,$\cdot$\,cm$^{-3}$),
\eq{ \frac{dT}{dr} < -(\mbox{0.4$-$0.8})\,
\mbox{MeV\,$\cdot$\,km$^{-1}$}\, ; \label{nch}}

\item for the transition between homogeneous matter and cheesed
phase ($\rho_{m} \approx 1.6 \times
10^{14}$~g\,$\cdot$\,cm$^{-3}$), \eq{ \frac{dT}{dr} >
-(\mbox{1.0$-$1.5})\, \mbox{MeV\,$\cdot$\,km$^{-1}$} \, .
\label{hch}}
\end{itemize}

The difference in the inequality signs in estimates (\ref{nch}) and (\ref{hch})
is connected with  the difference in the signs of the coefficients of the
temperature gradients in (\ref{main}) or (\ref{main1}) for the two phase
transitions under consideration.  Note that conditions (\ref{nch}) and
(\ref{hch}) are overlapping and complementary, so it is very likely that one of
them is satisfied. We, therefore, can expect that boiling of nuclear matter can
take place inside the supernova core. However, this only demonstrates the
possibility of principle, calling for additional thorough investigation of this
issue.

\section{A model}\label{Model}

In this section, we construct a simplified model of phase transition between a
heavier phase 1 and a lighter phase 2, which, for definiteness, we take to be
the cheesed phase and the phase of nuclei, respectively.

We assume spherical symmetry of the PNS, so that the cheesed phase is located
in a spherical shell with radial coordinate from $R - H$ to $R$. The free-fall
acceleration on the upper boundary of this region is equal to
\eq{g=\frac{GM}{R^{2}}=1.33\times 10^{14}
\left(\frac{M}{M_{\odot}}\right)\left(\frac{10 \, \mbox{km}}{R}\right)^{2}
\mbox{cm\,$\cdot$\,sec$^{-2}$} \, ,} where $M$ is the total mass inside the
sphere of radius $R$.

We can estimate $H$ using the condition of hydrostatic equilibrium:
\begin{equation}
\rho g H \approx \Delta p\, ,
\end{equation}
where $\Delta p$ is the dimension of the pressure interval in which the cheesed
phase exists. According to Table~\ref{input}, this pressure interval is
approximately equal to 0.83~MeV$\,\cdot$\,fm$^{-3}$.  We thus have
\begin{equation}
H \approx \frac{\Delta p}{\rho g} \approx 1.0 \times \left( \frac{\Delta
p}{0.83\, \mbox{MeV\,$\cdot$\,fm$^{-3}$}}\right)\left(\frac{10^{14}\,
\mbox{g\,$\cdot$\,cm$^{-3}$}}{\rho}\right)\left(\frac{1.33 \times 10^{14}\,
\mbox{cm\,$\cdot$\,sec$^{-2}$}}{g}\right)\, \mbox{km} \, .
\end{equation}

If we assume that boiling takes place in the whole volume of phase 1, then the
total mass of the boiling matter can be estimated as\footnote{Recent
simulations \cite{Sonoda07} show that the total mass of different phases
between nuclei and homogeneous nuclear matter (collectively called ``pasta
phases'' in \cite{Sonoda07}) inside the supernova just {\em before\/} the
bounce can amount to $0.13$$-$$0.30\, M_{\odot}$. This estimate somewhat
differs from (\ref{Mb-estim}) because, before the bounce, one deals with {\em
small\/} pressure gradients in an (almost) free-falling matter, while, after
the bounce, we have matter close to hydrostatic equilibrium with its large
pressure gradients. It is possible, in principle, that the
\textit{pre-bounce\/} boiling, if it occurs, can affect the propagation of
prompt shock wave, but we do not consider this issue here.} \eq{M_{b} \approx
4\pi\rho R^{2}H \lesssim 6.1 \times 10^{-2} \left(\frac{\Delta p}{0.83\,
\mbox{MeV\,$\cdot$\,fm$^{-3}$}}\right) M\, . \label{Mb-estim}}

The densities of phases 1 and 2 are related by $\rho_{1}$ and $\rho_{2} =
\rho_{1}(1 - \epsilon)$, where $\epsilon \ll 1$. According to \cite{OS82},
$\epsilon$ is equal to $0.1$ for the phase transition between the cheesed phase
and homogeneous matter phase, and to $0.2$ for the transition between nuclei
and cheesed phase. Lower estimates for $\epsilon$ are present in
\cite{suraud85}: it equals to $0.07$$-$$0.08$ for both phase transitions if the
entropy per baryon (which is conserved in the simulations) $S/A =  1.0$; for
$S/A = 1.5$, the value $\epsilon = 0.05$ is obtained for the transition between
nuclei and cheesed phase, and $\epsilon = 0.01$ for the transition between
cheesed phase and homogeneous matter. To be conservative, we use the lowest
estimate in this paper, namely,  $\epsilon = 0.01$.

The maxi\-mum acceleration which can be reached by a raising bubble (neglecting
the liquid resistance) is \eq{a_{\rm max} = \frac{\epsilon GM}{R^{2}}\, .} The
maximum velocity that can be reached by the bubble is then \eq{v_{\rm max} \sim
\left( \frac{2\epsilon \Delta p}{\rho} \right)^{1/2} = 5100 \sqrt{\left(
\frac{\epsilon}{10^{-2}} \right) \left( \frac{\Delta
p}{0.83\,\mbox{Mev\,$\cdot$\,fm$^{-3}$}} \right)\left(\frac{10^{14}\, \mbox{g
$\cdot$cm$^{-3}$}}{\rho}\right)} \, \mbox{km\,$\cdot$\,sec$^{-1}$} \, .}

Matter in the boiling volume will convectively move in both directions. We can
expect that roughly half of matter moves upwards (the bubbles and the
surrounding matter) and the other half moves downwards with the same average
velocity (from the momentum conservation). Therefore, the ``convective
boiling'' overturn will be established. We can try to estimate the maximum
efficiency of this overturn.

The first effect to be discussed is heat transfer. If the bubbles fill half of
the boiling volume, the heating rate at the surface of the boiling layer is
\begin{eqnarray} \label{heating}
\dot{Q}_{\rm max} &\sim& \frac{1}{2}\times 4\pi R^{2} v_{\rm max}\,q
\nonumber \\
&=& 7.7 \times 10^{52}\, \left(\frac{R}{10\, \mbox{km}}
\right)^{2}\left(\frac{v_{\rm max}}{5100\,
\mbox{km\,$\cdot$\,sec$^{-1}$}}\right)\left(\frac{q}{0.015\,
\mbox{MeV\,$\cdot$fm$^{-3}$}}\right)\,\mbox{erg\,$\cdot$\,sec$^{-1}$}\, .
\end{eqnarray}
Here, $q$ is the specific volume heat of evaporation, and its value is
estimated as $q=0.015$\,MeV\,$\cdot$\,fm$^{-3}$ from \cite{RPW83}. The value
(\ref{heating}) is comparable to the neutrino luminosity. This is, in
principle, the upper estimate of heating rate which corresponds to maximal
workload of the ``boiling machine.'' In reality, the boiling heat transfer
should work together with diffusion and/or convection below and above the
boiling shell, increasing the {\em net\/} heat transfer rate.\footnote{This can
be compared to an electric circuit with a series of resistances. As one of the
resistances in the series is shortened out, the total current increases.} This
should lead to an increase in pressure behind the shock, providing more
efficient conditions for shock revival.

The second effect is the momentum transfer. The maximum mechanical pressure
that the bubbles can exert is estimated as \eq{p_{\rm max} \sim \rho v_{\rm
max}^{2} = 2\epsilon \Delta p = 1.7 \times 10^{-2}\,
\mbox{MeV\,$\cdot$\,fm$^{-3}$}\, ,} where the numerical value corresponds to
the cheesed phase. This is much smaller than {\bf $\Delta p$} since $\epsilon
\ll 1$; therefore, including this contribution to pressure does not seriously
affect our previous calculation. But it can provide an additional barrier for
the infalling matter to reach the inner core, reducing the final mass of the
neutron star.

We should admit that most of the above estimates represent upper limits
corresponding to the maximal workload of the ``boiling machine.'' In a
subsequent paper, we will discuss these processes in greater detail.

\section{Summary and conclusions}\label{Discussion}

In this paper, we have demonstrated the possibility of boiling of nuclear
liquid in the supernova core after the bounce. If it occurs, it can lead to
effects increasing the efficiency of the neutrino-driven mechanism of supernova
explosions. Among these effects are the following:
\begin{itemize}
\item The increase of heat transfer rate from the inner core to the neutrinosphere. This
increases the mean neutrino energy, making the delayed-shock mechanism more
efficient.

\item The increase of pressure in the boiling region. It provides an
additional barrier between the infalling matter and the inner core, thereby
reducing the mass of the neutron-star remnant.

\end{itemize}

We expect that taking into account the effect of boiling in the conventional
delayed-shock/acoustic mechanism is important and will enable one to explain
the energetics of the supernova explosions in a more simple and self-consistent
way.

\section*{Acknowledgements}

D.~I.\@ is grateful to the Scientific and Educational
Center\footnote{http://sec.bitp.kiev.ua} of the Bogolyubov Institute for
Theoretical Physics in Kiev, Ukraine, and especially to Dr.~Vitaly Shadura for
creating wonderful atmosphere for young scientists, and to T.~Foglizzo,
M.~Liebend\"{o}rfer and D.~K.~Nadyozhin for their helpful comments. This work
was supported in part by grant No.~5-20 of the ``Cosmomicrophysics'' programme
of the Ukrainian Academy of Sciences and by the INTAS grant
No.~05-1000008-7865.

\end{document}